\documentclass[twocolumn,showpacs,preprintnumbers,nofootinbib,superscriptaddress,10pt]{revtex4-1}
\usepackage[utf8]{inputenc}
\usepackage{graphicx,amsmath,amssymb,xcolor,hyperref}
\usepackage{url}
\usepackage{lmodern}
 
\begin{document}

\preprint{MITP/19-061} 

\title{Discovering the $h\to Z \gamma$ Decay in $t \bar t$
Associated Production}

\author{Florian Goertz}
\email{florian.goertz@mpi-hd.mpg.de}
\affiliation{Max-Planck-Institut f\"ur Kernphysik, Saupfercheckweg 1, 69117 Heidelberg, Germany}
\author{Eric Madge}
\email{eric.madge@uni-mainz.de}
\affiliation{PRISMA$^+$  Cluster of Excellence and Mainz Institute for Theoretical Physics,\\Johannes  Gutenberg-Universit\"at  Mainz, 55099 Mainz, Germany}
\author{Pedro Schwaller}
\email{pedro.schwaller@uni-mainz.de}
\affiliation{PRISMA$^+$  Cluster of Excellence and Mainz Institute for Theoretical Physics,\\Johannes  Gutenberg-Universit\"at  Mainz, 55099 Mainz, Germany}
\author{Valentin Titus Tenorth}
\email{valentin.tenorth@mpi-hd.mpg.de}
\affiliation{Max-Planck-Institut f\"ur Kernphysik, Saupfercheckweg 1, 69117 Heidelberg, Germany}

\date{\today}

\pacs{
}

\begin{abstract}
We explore the prospects to discover the $h \to Z \gamma$ decay in $t\bar t$-associated production, featuring a signal-to-background ratio of ${\cal O}(1)$. Performing a detailed analysis of the semi-leptonic $t \bar t $-decay channel, we demonstrate that the production mode could lead to a $\sim5\,\sigma$ discovery at the high-luminosity LHC, while the effective $h Z \gamma$ coupling could be extracted with a $\sim15\,\%$ accuracy. Extending the analysis to potential future $pp$ colliders with 27\,TeV and 100\,TeV center-of-mass energies, we also show that the latter would allow precision measurements at the few percent level, rendering possible precise extractions of the spin and $CP$ properties of the Higgs boson.
\end{abstract}

\maketitle

\section{Introduction}

The decay of the Higgs boson to a photon and a weak $Z$ boson, $h \to Z \gamma$, has not been discovered yet. Measuring it can not only provide a further consistency test of the Standard Model (SM) of particle physics, but also has the potential to unveil new physics (NP) that could be hidden in other observables~\cite{Low:2011gn,Coleppa:2012eh,Azatov:2013ura,Contino:2013kra,Pomarol:2013zra,Elias-Miro:2013mua,Belanger:2014roa,Chen:2014ona,Arina:2014xya,Liu:2018qtb}.
Moreover, in principle it furnishes a promising channel to extract spin and parity properties of the Higgs boson.

The decay is challenging to access via production modes entertained so far, such as $g g \to h$, which lead to an expected significance of $2\,\sigma$ with 100~fb$^{-1}$ at the 14 TeV LHC~\cite{Gainer:2011aa}. 
Refined projections by the ATLAS and CMS collaborations show that even at the end of the LHC programme, with 3~ab$^{-1}$, a $5\,\sigma$ discovery will be challenging~\cite{ATL-PHYS-PUB-2018-054}.
Latest experimental searches using $139\,$fb$^{-1}$ of data set an upper limit of $3.6$ times the SM value for $\sigma(pp \to h\to Z\gamma)$ \cite{Aad:2020plj,Aaboud:2017uhw, Sirunyan:2018tbk}.

The $h \to \ell^+ \ell^- \gamma$ channel also offers the possibility to independently measure the spin~\cite{Freitas:2010ht,Gainer:2011aa} and $CP$~\cite{Chen:2014ona} properties of the Higgs, but the low signal to background ratio makes it difficult to extract angular correlations or asymmetries in the inclusive search. Here and in the following $\ell$ always denotes electrons and muons.

In this article, we entertain the channel $p p \to t \bar t h$, $h\to Z \gamma \to \ell^+ \ell^- \gamma$, which enhances the prospects to discover the $h \to Z \gamma$ decay and to measure the corresponding effective coupling. In fact, the $t \bar t h$ production mode has recently been observed by ATLAS and CMS, inviting to use it for further studies \cite{Aaboud:2018urx, Sirunyan:2018hoz, ATLAS:2019aqa}.
It profits in particular from the large Yukawa coupling of the top quark, such that the radiation of a Higgs boson from a $t \bar t$ state leads only to a modest suppression of the cross section.
This promises a significantly enlarged signal-to-background ratio compared to other channels like gluon fusion, where one starts inevitably with a further loop-suppressed signal, thereby increasing the prospects to measure spin and $CP$.
We will both study the expected significance for the channel under consideration at the high-luminosity LHC (HL-LHC) as well as examine potential constraints on the coefficient of the effective $h Z \gamma$ coupling.
Finally, we will extend the analysis to include a future 27\,TeV~(HE-LHC) and a 100\,TeV $pp$ collider, like the FCC$_{hh}$.

\section{Setup}

We consider the SM, augmented with the $D=6$ operators 
\begin{equation}
\label{eq:ops}
\begin{split}
	{\cal O}_{HW}=&\frac{i g}{m_W^2} \left( D^\mu H \right)^\dagger 
	\sigma_i \left( D^\nu H \right) W_{\mu\nu}^i \,,\\
	{\cal O}_{HB}=&\frac{i g^\prime}{m_W^2} \left( D^\mu H \right)^\dagger 
	\left( D^\nu H \right) B_{\mu\nu}\,,\\
	{\cal O}_\gamma=&\frac{{g^\prime}^2}{m_W^2} |H|^2
	B_{\mu\nu}B^{\mu\nu} \,,
\end{split}
\end{equation}
relevant for the decay $h \to Z\gamma$ to leading approximation\footnote{Thus, we do not entertain possible NP effects in Higgs {\it production}.
Furthermore, we neglect $CP$ odd operators.}, where $H$ is the scalar Higgs doublet, parametrised after electroweak symmetry breaking (EWSB) as \mbox{$H= 1/\sqrt 2 (- i \varphi_1 - \varphi_2,\, v + h + i \varphi_3)^T$}. 
Here, $v$ denotes the vacuum expectation value (VEV) of the Higgs field $\langle H \rangle = 1/\sqrt 2 (0,v)^T$, which triggers EWSB, $h$ is the physical radial Higgs boson and $\varphi_{1,2,3}$ are the Goldstone modes. This setup allows us to study deviations from the SM in a model independent way, under the assumption that there is a mass gap between the SM and the NP.
After EWSB, the operators~(\ref{eq:ops}) generate in particular the Lagrangian term
\begin{equation}
	{\cal L \supset}\ c_{Z \gamma} \frac h v Z_{\mu\nu} \gamma^{\mu\nu},
\end{equation}
at the tree-level, contributing to the $h \to Z \gamma$ decay, with
\begin{equation}
	\label{eq:cZa}
	c_{Z\gamma}=- \tan \theta_W \left[ (c_{HW}-c_{HB}) 
	+ 8 \sin^2\theta_W c_\gamma \right]\,,
\end{equation}
where $c_{HW,HB,\gamma}$ are the coefficients of the operators \eqref{eq:ops} in the effective $D=6$ Lagrangian. 
Note that the direction \eqref{eq:cZa} is not very constrained yet such that still significant NP effects can be present \cite{Pomarol:2013zra, deBlas:2018tjm, Ellis:2018gqa, Biekotter:2018rhp}.

For the following analysis we define the ratio of the decay width in the presence
of the operators \eqref{eq:ops} and the SM decay width (see, e.g., \cite{Contino:2013kra})
\begin{equation}
	\label{eq:GamFrac}
	\frac{\Gamma(h \to Z \gamma)}{\Gamma(h \to Z \gamma)_{\rm SM}} \equiv \kappa_{Z\gamma}^2
	\simeq 1 - 0.146 \frac{4 \pi}{\alpha \cos \theta_W} c_{Z\gamma}\,, 
\end{equation}
where the second equality is valid for small $c_{Z\gamma}$.
We will eventually study the constraints that can be set on $\kappa_{Z\gamma}$, and thus on the Wilson coefficient $c_{Z\gamma}$, from the process under consideration.

\section{Estimate}

In the SM, the cross section for producing a Higgs boson in association with two top quarks at the 14~TeV LHC including NLO QCD+EWK corrections is $\sigma(p p\to t \bar t h) = 613\,\text{fb}\ \substack{+6.0\%\\-9.2\%}\,(\text{scale})\, \pm 3.5\%\,(\text{PDF}+\alpha_s)$,\linebreak  while the relevant branching ratio amounts to ${{\cal B}(h \to Z \gamma)= 1.54\cdot 10^{-3}}$~\cite{deFlorian:2016spz}. 
We consider the $Z$ boson decaying to two leptons, $\ell=e,\mu$, which has a branching fraction of ${\cal B}(Z \to \ell^+ \ell^-)= 2 \times 0.0336 = 0.067$~\cite{Tanabashi:2018oca}. For the HL-LHC with $3\,$ab$^{-1}$ of integrated luminosity we thus expect $S_0 \approx 190$ signal events over all top decay channels.

For the signal to remain observable after selection cuts, the analysis will have to be as inclusive as possible.
Electrons, muons, and photons are reconstructed with high efficiencies. On the other hand, tagging $t\bar{t}$-associated production and including isolation requirements, taking into account the probability of overlapping with some of the top decay products, will reduce the number of events. For a first estimate, we thus assume a selection efficiency of $(10-15)\,\%$, comparable to the experimental efficiency of the di-photon channel \cite{TheATLAScollaboration:2013mia}, which we will corroborate quantitatively in an explicit analysis of the semi-leptonic top-decay channel in the next section. This would finally lead to about $S=(20-30)$ signal events per experiment.

The main irreducible background is $t\bar{t} Z$ production with radiation of a photon from initial or final states. At the 14~TeV LHC, the NLO QCD cross section  with ${p_{T,\gamma} > 10\,\text{GeV}}$ and ${|\eta_\gamma| < 4.0}$ is $\sigma(pp \to t\bar{t} Z \gamma)=9.3\,\text{fb}$, about ten times larger than the signal, resulting in $B_0 \approx 1870$.

Among the reducible backgrounds, we expect the dominant contribution from $pp \to t jj Z\gamma$ and $pp \to t\bar{t} Z j$ production, where $j$ denotes a jet in the 5-flavor scheme including $b$-jets. The former background is only relevant when considering the semi-leptonic and fully-hadronic channels; and in the latter case one jet is misidentified as a photon. Experimentally, the latter background can be estimated by loosening the photon identification, however we cannot simulate this reliably.
Eventually the best approach will be to float the background normalisation to fit the data in the side-bands below and above $m_h$. For the purpose of the present estimate we account for reducible backgrounds by simply increasing the irreducible background cross section by $50~\%$ to obtain more realistic estimates for the sensitivity. Including this factor and multiplying with the selection efficiency above, assuming that the efficiencies for signal and background are comparable if no cut is applied to the $Z\gamma$ invariant mass, we arrive at $(280-420)$ background events.
Whether other backgrounds are relevant will depend on the $t\bar{t}$ decay channel and on the analysis, but we expect them to be sub-leading and have a smooth $m_{\gamma\ell\ell}$ invariant mass distribution.

Once the $\gamma \ell^+ \ell^-$ invariant mass is restricted to a $10$~GeV window around the Higgs mass, the background is reduced by another factor of $\sim 15$, see below, and we would obtain $B=(20-30) \approx S$, resulting in a $4.5\,\sigma-5.5\,\sigma$ sensitivity from a simple cut and count analysis.
This can further be improved by fitting the invariant mass distribution with signal plus background and background only hypotheses. The potential to observe the $h\to \ell^+ \ell^- \gamma$ channel in a low background environment is our main motivation to perform this study.
In the next section we provide a detailed simulation for the semi-leptonic $t\bar{t}$ channel, to better understand how realistic the above estimate is.

\section{Analysis}

To get a solid estimate of the expected sensitivity at the HL-LHC with $\sqrt{s}=14$~TeV and an integrated luminosity of 3~ab$^{-1}$ we here perform an analysis of the semi-leptonic channel using  Monte Carlo simulation, and then use the resulting selection efficiency to estimate the sensitivity including all top-pair decay channels in section~\ref{sec:sensitivity}. 
We simulate the signal process $pp \to t \bar{t}h$ with MadGraph5\_aMC@NLO~\cite{Alwall:2014hca, Hirschi:2015iia} at next-to-leading order (NLO) in QCD using the PDF4LHC15\_nlo\_30\_pdfas PDF set~\cite{Butterworth:2015oua} provided through LHAPDF6~\cite{Buckley:2014ana}.
Our value for the $t \bar{t} h$-production cross-section is in good agreement with the results of the LHCHXSWG, quoted above.
For the parton-showering we use the MadGraph-build-in Pythia~$8.2$~\cite{Sjostrand:2014zea}, only allowing for the $h\to Z\gamma$ and ${Z \to \ell^+ \ell^-}$ decays and rescaling the cross-section by the branching fractions ${{\cal B}(h \to Z \gamma)}$ and ${{\cal B}(Z \to \ell^+\ell^-)}$. A fast detector simulation is done with Delphes $3.4.2$~\cite{deFavereau:2013fsa} using the HL-LHC detector card.

We also simulate several background processes. The most relevant ones are, the irreducible background ${pp \to t \bar{t} \gamma Z\,,\,Z\to \ell^+\ell^-}$, without contributions from Higgs decays giving a cross section of approximately $620$~ab at NLO in QCD for $p_{T,\gamma} > 10$~GeV and $|\eta_\gamma| < 4$, and the reducible one $pp\to t j j \gamma Z\,,\,Z\to \ell^+\ell^-$ with a LO cross section of $940$~ab. 
We do not simulate the $t\bar{t} Z j$ background, as we cannot model the jet misidentification reliably.
Instead, it is accounted for by enhancing the $t\bar{t} Z \gamma$ background in our calculation of the significance, expecting the $t\bar{t} Z j$ background to amount to roughly 20~\% of the $t\bar{t} Z \gamma$ one~\cite{Aaboud:2017uhw}.
Other possible final states, such as $t\bar t j W^\pm \gamma$, $W^\pm b\bar{b} j Z \gamma$ and $t\bar{t} t\bar{t}\gamma$ have negligible cross sections in the selected region and sum up to less than 10\% of the total background events.

We focus on semi-leptonic $t\bar{t}$ decays ($t \!\to\! b j j,\ \bar t\! \to\! \bar b \ell^- \bar \nu_\ell$, or vice versa) as those are best to handle for a cut-and-count analysis and comment on the hadronic and leptonic channel in the next section. Still all top decays are allowed in Pythia to account for example for the possibility of $\tau$'s being mistagged as leptons and therefore contributing to the semi-leptonic channel.

The reconstruction requirements for electrons (muons) in Delphes are $p_T > 15~(10)$~GeV, $|\eta| < 2.47~(2.7)$ and for photons $p_T > 5$~GeV, $|\eta| < 2.37$, and it is demanded to have no selected leptons within a cone of $R=0.3$.
Jets are reconstructed with FastJet~3~\cite{Cacciari:2011ma} using the anti-$k_t$ algorithm~\cite{Cacciari:2008gp} with ${R=0.4}$ and are considered to have $p_{T,j}>25$~GeV and $|\eta|<2.5$. 
In addition the following selection requirements motivated by experimental analyses \cite{Aaboud:2017uhw, TheATLAScollaboration:2013mia} have to be fulfilled\footnote{Note that these cuts are mainly meant to select/specify our signal and suppress other backgrounds rather than to separate it from the irreducible background.}:

\begin{itemize}
	\item Exactly three leptons (electrons and muons) satisfying the reconstruction requirements
	\item Three or more jets
	\item $p_{T,j}>30$~GeV for the first three jets
	\item Missing energy $E\!\!\!/_{T}>20$~GeV 
	\item At least one $b$-tagged jet 
	\item At least one photon with $p_{T,\gamma}>15$~GeV 
	\item $Z$-reconstruction: OSSF lepton pair with 76~GeV $ \!<\! m_{\ell\ell} \!<\!$ 106~GeV
	\item Higgs-reconstruction: 120~GeV $\!<\! m_{\gamma\ell\ell} \!<\!$ 130~GeV
\end{itemize}

\begin{figure}
	\includegraphics[width=.45\textwidth]{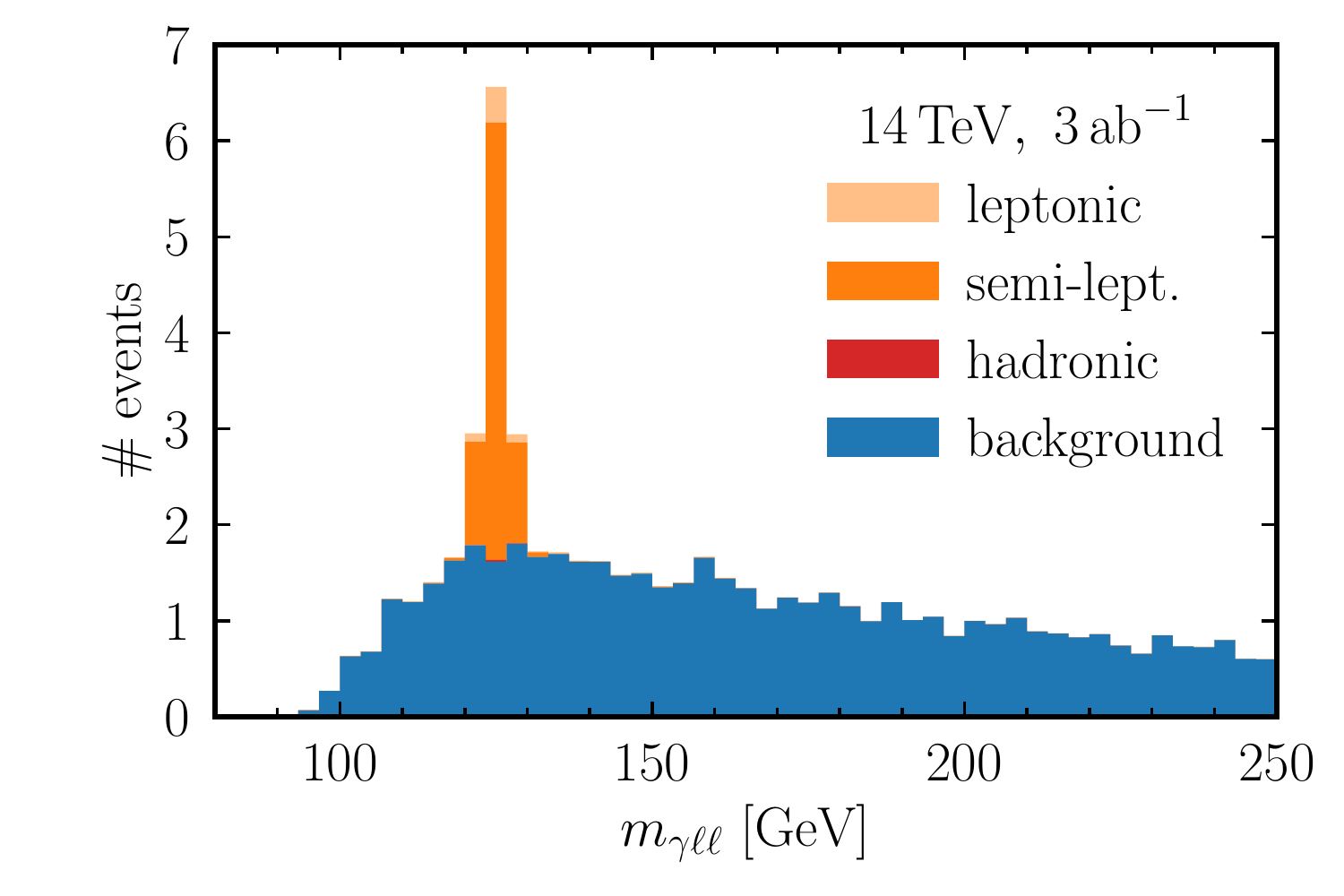}
	\caption{The invariant mass spectrum for the signal process, stacked on the irreducible background distribution (blue), before Higgs-reconstruction cut, for the top-quark pair decaying hadronically (red, not visible), semi-leptonically (orange) or leptonically (light orange).}
	\label{fig:histo14}
\end{figure}

To reconstruct the $Z$-boson we require an opposite sign, same flavour (OSSF) lepton pair in the invariant mass range ${76\,\text{GeV} \!<\! m_{\ell\ell} \!<\! 106\,\text{GeV}}$ in the final state, avoiding contamination from top-decays. If more than one lepton pair fulfils this requirement, the one closer to the $Z$-mass is chosen. This lepton pair together with the highest-$p_T$ photon is used to reconstruct the Higgs mass.
The invariant mass distribution of the $\gamma \ell^+\ell^-$ system (before applying the $m_{\gamma\ell\ell}$ cut) is shown in Fig.~\ref{fig:histo14}.

The numerical results for the signal and the two backgrounds are shown in Table~\ref{tab:cutflow14}.
The signal clearly peaks at $m_{\gamma\ell\ell} = m_h=125$~GeV and we see that by cutting on a window of $m_h \pm 5$~GeV, which is experimentally feasible \cite{Aaboud:2017uhw, Sirunyan:2018tbk}, we can obtain $S/B \gtrsim 1$.

The signal and background selection efficiencies for the semi-leptonic channel follow from Table~\ref{tab:cutflow14} as ${\epsilon_N \equiv N_{\rm final}/(N_{\rm initial}\,{\cal B}_{\rm semi-lept.}),\, N=S,\, B_{\rm irred},\, B_{\rm red}}$, where ${\cal B}_{\rm semi-lept.} = {\cal B}_{t\bar{t}\to b\bar{b}\ell\nu jj} =0.288$ for the signal and irreducible background, and ${\cal B}_{\rm semi-lept.} = {\cal B}_{t\to b\ell\nu} = 0.213$ for the $tjjZ\gamma$ background~\cite{Tanabashi:2018oca}.
We obtain $\epsilon_S=0.14$, $\epsilon_{B_{\rm irred}}=0.0097$ and $\epsilon_{B_{\rm red}}=0.0027$. As to expect, the reducible background has a smaller selection efficiency than the irreducible one.\footnote{Before Higgs-reconstruction, we get $\epsilon_{B_{\rm irred}}=0.15 \approx \epsilon_S$.} 

\begin{table}[!t]
	\begin{tabular}{l||c|c|c}
	Cut & $\quad\; S\quad\; $ & \; $ttZ\gamma$ \quad & \; $tjjZ\gamma$\quad \\\hline
	Initial & 186 & 1862 & 2817 \\
	$N(l)=3$ & 25 & 273 & 209 \\
	$N(j) \geq 3,\, p_{T,j}\!>\! 30$~GeV & 15 & 170 & 46 \\
	$E\!\!\!/_{T}>20$~GeV  & 14 & 160 & 41 \\
	$N(b)\geq  1$ & 12 & 137 & 34 \\
	$ N(\gamma)\geq 1,\, p_{T,\gamma}\!>\! 15$~GeV & 8.1 & 83 & 21 \\
	$Z$-reconstruction & 7.6 & 80 & 21 \\
	Higgs-reconstruction & 7.3 & 5.2 & 1.6
	\end{tabular}
	\caption{Signal $S$ and background events for two main processes $t\bar t\gamma Z(\to \ell^+\ell^-)$ and $tjj\gamma Z(\to \ell^+\ell^-)$ after each requirement to select the semi-leptonic channel for the HL-LHC with $\sqrt{s}=14\,\text{TeV}$ and $3\,\text{ab}^{-1}$.  For the backgrounds, a cut of $p_{T,\gamma} >10$~GeV and $|\eta_\gamma| < 4$ is imposed at the generator level.}
	\label{tab:cutflow14}
\end{table}

\pagebreak

\section{Sensitivity estimate}
\label{sec:sensitivity}

In order to arrive at our final result for the expected significance and the anticipated constraint on $\kappa_{Z\gamma}$ we assume that the efficiencies derived above for the semi-leptonic top-decay channel hold also for the leptonic and hadronic channels.

The reducible $pp\to tjjZ\gamma$ background is specific to the semi-leptonic and fully-hadronic channel. We therefore do not use the result of its simulation directly, but include it in the rescaling of the irreducible background.
From the proper simulation of the process in the semi-leptonic channel we find that the number of background events is increased by approximately $30~\%$ compared to the irreducible-background-only case.
In the following, we thus increase the irreducible background by  $50~\%$, also accounting for a $20~\%$ enhancement~\cite{Aaboud:2017uhw} from the $t \bar{t} Z j$ contribution.

We thus finally arrive at a total $S = 186 \times \epsilon_S \approx 25$ and $B = 1.5 \times 1862 \times \epsilon_{B_{\rm irred}} \approx 27$, including now realistic analysis cuts and taking into account losses due to overlapping final state particles in a detector simulation. This result agrees well with our first estimate above.
Considering the statistical error of $\Delta B = \sqrt{B} \approx 5$, we thus expect to establish a signal from the total rate alone with a significance $ S/
\sqrt{B} \approx 5\,\sigma$ at a single experiment.

A more precise definition of the discovery significance is given by $Z = \sqrt{2\left((S+B)\log (1+S/B) -S \right)}$~\cite{Cowan:2010js}, which converges to $S/\sqrt{B}$ for $S \ll B$. Employing this formula we find a more conservative significance of $Z=4.3$. We expect that the sensitivity can be further improved by performing a likelihood-fit of the signal over a smooth background, and thus a discovery should be feasible in this channel. 
As this would add further experimental uncertainties which can only be estimated using a full detector simulation, we decided to stay conservative and not use shape information here. 

Finally, the recent developments in top-reconstruction using boosted decision trees allow to identify hadronic top-decays with a high efficiency, which could further enhance the sensitivity. The selection efficiency is at least comparable to the leptonic channel in~\cite{CMS-PAS-HIG-18-018, Aaboud:2018urx, Sirunyan:2018hoz, ATLAS:2019aqa}, thus justifying our extrapolation from the semi-leptonic to the hadronic channel.

\section{27 and 100~TeV Colliders}

Next we study the channel under consideration at a future $27~$TeV ($100~$TeV) $pp$ collider with 15~ab$^{-1}$ (30~ab$^{-1}$) of integrated luminosity~\cite{Abada:2019ono,Benedikt:2018csr}.

Here the $t\bar t h$ production cross section amounts to $2.9$~pb for 27~TeV~\cite{Cepeda:2019klc} and approximately $33$~pb for 100~TeV center of mass energy~\cite{Contino:2016spe}, which were reproduced by our MadGraph simulations. The background of $t \bar t Z\gamma$ production features 46~fb (670~fb) at 27~TeV (100~TeV) with $p_{T,\gamma} > 10$~GeV and $|\eta_\gamma|<4$. 
For simplicity and easier comparability we use a similar setting and the same reconstruction and selection requirements as for the HL-LHC.
We note that these cuts are rather low for the higher center-of-mass energies, but a detailed study of future collider settings is beyond the scope of this article and a moderate increase in the cuts is expected to have only a mild influence on the obtained results.
For the 100~TeV case we use the FCC$_{hh}$ Delphes Card.

Considering again the $Z \to \ell^+\ell^-$ channel, we obtain the cut-flows shown in Table \ref{tab:cutflow27}. The corresponding $m_{\gamma\ell\ell}$ invariant mass spectra can be found in the Supplemental Material~(Fig.~\ref{fig:histo27}). 
For both scenarios the same extrapolation to include all top-decay channels and an enhancement of the background by $50\,\%$ as for the HL-LHC is performed, motivated by our previous findings.

\begin{table}[!t]
	\begin{tabular}{l||c|c||c|c}
	 & \multicolumn{2}{c||}{$27\,$TeV, $15\,$ab$^{-1}$} & \multicolumn{2}{c}{$100\,$TeV, $30\,$ab$^{-1}$} \\
	Cut  & $\quad\; S \quad\;$  & $B$ & $\quad\; S\quad\;$ & $B$ \\\hline
	Initial &  4.4k & 47k &  112k & 1.3M \\
	$N(l)=3$ & 539 & 6.2k & 16k & 210k \\
	$N(j) \geq 3,\, p_{T,j}\!>\! 30$ GeV & 344 & 4.1k & 12k & 160k \\
	$E\!\!\!/_{T}>20$~GeV  & 322 & 3.9k & 11k & 150k\\
	$N(b)\geq 1$ & 276 & 3.3k & 10k & 140k \\
	$ N(\gamma)\geq1,\, p_{T,\gamma}\!>\! 15$ GeV & 180 & 2.0k & 6.7k & 84k \\
	$Z$-reconstruction & 166 & 1.9k & 6.3k & 82k \\
	Higgs-reconstruction & 160 & 101 & 6.1k & 3.2k
	\end{tabular}
	\caption{Number of signal $S$ and background $B$ events after each of the selection requirements at a $27\,$TeV or $100\,$TeV collider, with $3\,$ab$^{-1}$ and $15\,$ab$^{-1}$ of luminosity, respectively. For the background, a cut of $p_{T,\gamma} >10\,$GeV and $|\eta_\gamma| < 4$ is imposed at the generator level. 
	}
	\label{tab:cutflow27}
\end{table}

\section{Constraints on $\kappa_{Z\gamma}$}

In the following, we want to examine the expected constraints that can be set on $\kappa_{Z \gamma}$ from the process under consideration.
To that end, we first calculate the predicted number of events $N(\kappa_{Z \gamma})= S(\kappa_{Z \gamma})+B$, where $S(\kappa_{Z \gamma})$ is obtained from the SM value $S=25$ by multiplying with $\kappa_{Z \gamma}^2$, see~(\ref{eq:GamFrac}).
We further assume the SM to be true and calculate how many standard deviations $\Delta N(\kappa_{Z \gamma})$ away the prediction $N(\kappa_{Z \gamma})$ is from $N(\kappa_{Z \gamma}=1)$, which is the expected outcome of the experiment. 
The values of $\kappa_{Z \gamma}$ that lead to a discrepancy of more than $n$ standard deviations are then expected to be excluded with a significance of $n\, \sigma$.

Following this procedure for the three considered colliders, the expected $1\sigma$ ($2\sigma$) constraints on $\kappa_{Z \gamma}$ are thus obtained as
\begin{align}
	\hspace*{-7pt}14\,\text{TeV}:\ \hphantom{0}0.86 &\leq \kappa_{Z \gamma} \leq 1.14 \ \hspace*{10pt}(0.71 \leq \kappa_{Z \gamma} \leq 1.29) \notag\\
	\hspace*{-7pt}27\,\text{TeV}:\ \hphantom{0}0.97 &\leq \kappa_{Z \gamma} \leq 1.03 \ \hspace*{10pt} (0.94 \leq \kappa_{Z \gamma} \leq 1.06) \\
	\hspace*{-7pt}100\,\text{TeV}:\  0.995 &\leq \kappa_{Z \gamma} \leq 1.005 \ (0.991 \leq \kappa_{Z \gamma} \leq 1.009)\,, \notag
\end{align}
and presented as red bars in Fig.~\ref{fig:kappa}. The corresponding $p$-value plots can be found in the Supplemental Material~(Fig.~\ref{fig:p}).

\begin{figure}
	\includegraphics[width=.45\textwidth]{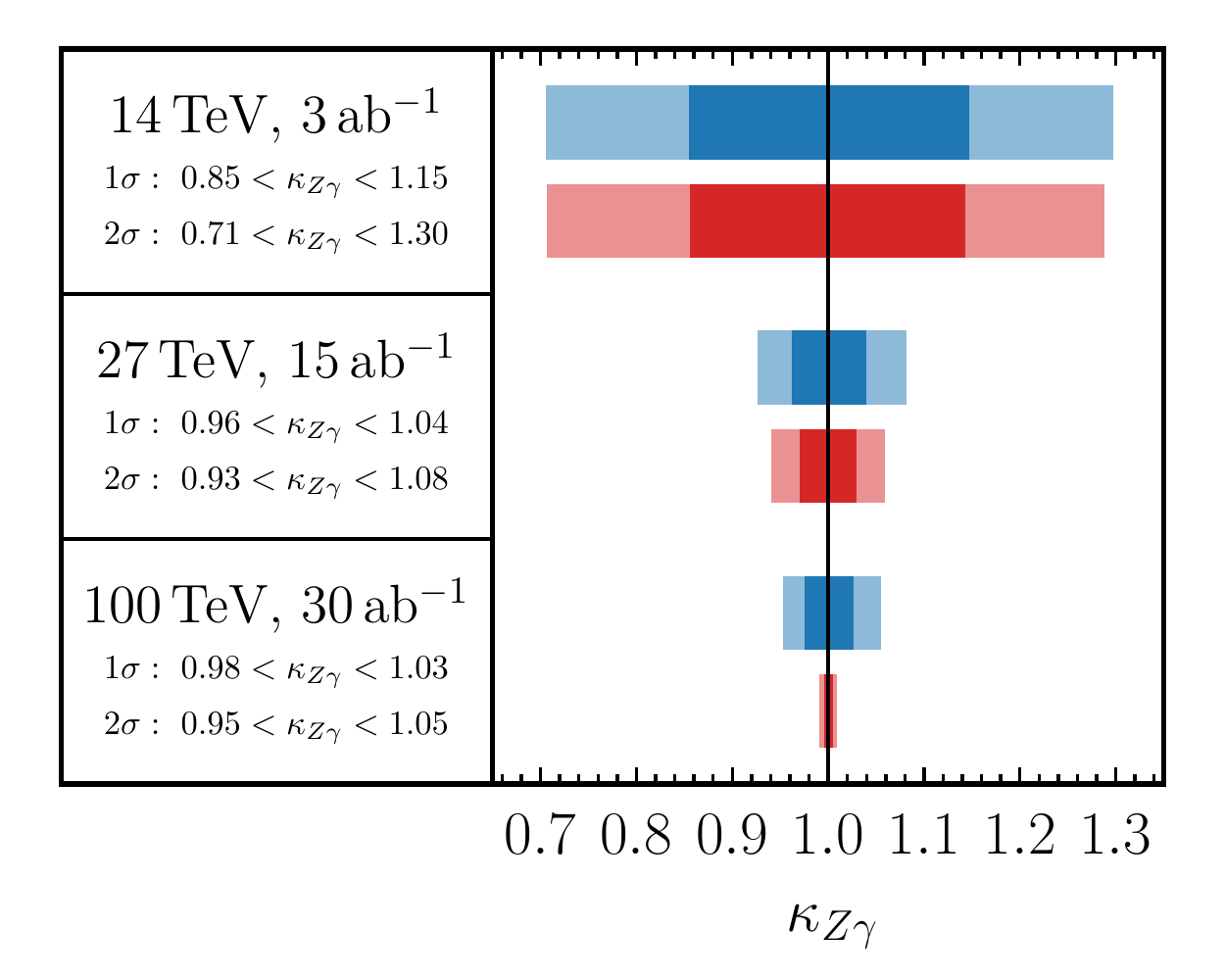}
	\caption{$1\,\sigma$ and $2\,\sigma$ limits on $\kappa_{Z\gamma}$ assuming the SM to be true, as obtained from our analysis. Shown are limits with statistical errors only (red) and including a $5\,\%$ systematic error from the theory uncertainty in the $t\bar{t}h$ cross section (blue). The numbers in the left column include the $5\,\%$ uncertainty.}
	\label{fig:kappa}
\end{figure}

At envisaged future hadron colliders, a signal in this low background process could thus be established at a level of well beyond $5\,\sigma$, where the number of events clearly allows to pin down quantities like the spin of the Higgs boson or its $CP$ properties and to perform precision tests of the effective $h Z \gamma$ coupling at the $1\,\%$ level. 

At this level of precision, it becomes necessary to take into account potential systematic errors.
On the experimental side there are $\mathcal{O}(1-5\,\%)$ uncertainties related to the lepton, photon and $b$-jet identification, which could be further reduced by fitting the sidebands of the spectra. Nevertheless a full experimental analysis is needed to assess these uncertainties and established the estimated precision.
On the theory side, the interpretation of the observed rate as a constraint on $\kappa_{Z\gamma}$ is affected by the uncertainty in $\sigma (pp \to t\bar{t}h)$, which is currently of order $10\,\%$ for the LHC. Anticipating some theory progress, in Fig.~\ref{fig:kappa} we show in addition the level of precision that is obtained assuming a $5\,\%$ systematic error (blue bars). 
The projected $1\,\sigma\ (2\,\sigma)$ constraints then become
\begin{align}
	14\,\text{TeV}:\ \ & 0.85 \leq \kappa_{Z \gamma} \leq 1.15 \quad (0.71 \leq \kappa_{Z \gamma} \leq 1.30) \notag\\
	27\,\text{TeV}:\ \ & 0.96 \leq \kappa_{Z \gamma} \leq 1.04 \quad (0.93 \leq \kappa_{Z \gamma} \leq 1.08) \\
	100\,\text{TeV}:\ \ & 0.98 \leq \kappa_{Z \gamma} \leq 1.03 \quad (0.95 \leq \kappa_{Z \gamma} \leq 1.05)\,. \notag
\end{align}
Our projected sensitivities to $\kappa_{Z\gamma}$ are comparable to those in other Higgs production channels, which are on the order of 10\,\% (3 -- 4\,\%) at the HL-(HE-)LHC~\cite{Cepeda:2019klc}.

A further reduction of systematic errors could be achieved if one considers ratios of couplings such as $\kappa_{Z\gamma}/\kappa_{\gamma\gamma}$ in the $t\bar{t}h$ channel. Such ratios are very sensitive to potential new physics patterns, for example additional charged fermions coupled to the Higgs have a stronger effect on $\kappa_{\gamma\gamma}$, since the contribution of the $W$ boson loop strongly dominates the $h\to Z\gamma$ rate in the SM.

\section{Conclusions}
\label{sec:conclusions}

We have explored the prospects to discover the decay of the Higgs boson to a photon and a $Z$ boson in $t \bar t $-associated Higgs production. Focusing our analysis on the semi-leptonic $t \bar t $-decay channel, we demonstrated that the production mode considered could lead to a $\sim5\,\sigma$ discovery at the HL-LHC.
Beyond that, we derived projected bounds on the effective $h Z \gamma$ coupling, $\kappa_{Z \gamma}$, at the HL-LHC and future $pp$ colliders with $27\,$TeV and $100\,$TeV center-of-mass energies, finding
$1\,\sigma$ constraints at the level of $15\,\%$, $4\,\%$, and $2\,\%$, respectively. The sensitivity is comparable to or even exceeds that of future lepton colliders~\cite{Cao:2015iua,No:2016ezr,Durieux:2017rsg}.
Finally, the corresponding $S/B$ ratios of ${\cal O}(1)$ would also render possible precise extractions of the spin and $CP$ properties of the Higgs boson.

\acknowledgements{
We are grateful to Alex Azatov for useful discussions. 
Research in Mainz is supported by the Cluster of Excellence ``Precision Physics, Fundamental Interactions, and Structure of Matter" (PRISMA+ EXC 2118/1) funded by the German Research Foundation(DFG) within the German Excellence Strategy (Project ID 39083149), and by grant 05H18UMCA1 of the German Federal Ministry for Education and Research (BMBF).
The authors gratefully acknowledge the computing time granted on the supercomputer Mogon at Johannes Gutenberg University Mainz (\url{hpc.uni-mainz.de}).
VT acknowledges support by the International Max Planck Research School for Precision Tests of Fundamental Symmetries.
}

\bibliography{refs.bib}

\newpage
\onecolumngrid
\section*{Supplemental material}

\begin{figure*}[h]
	\includegraphics[width=.45\textwidth]{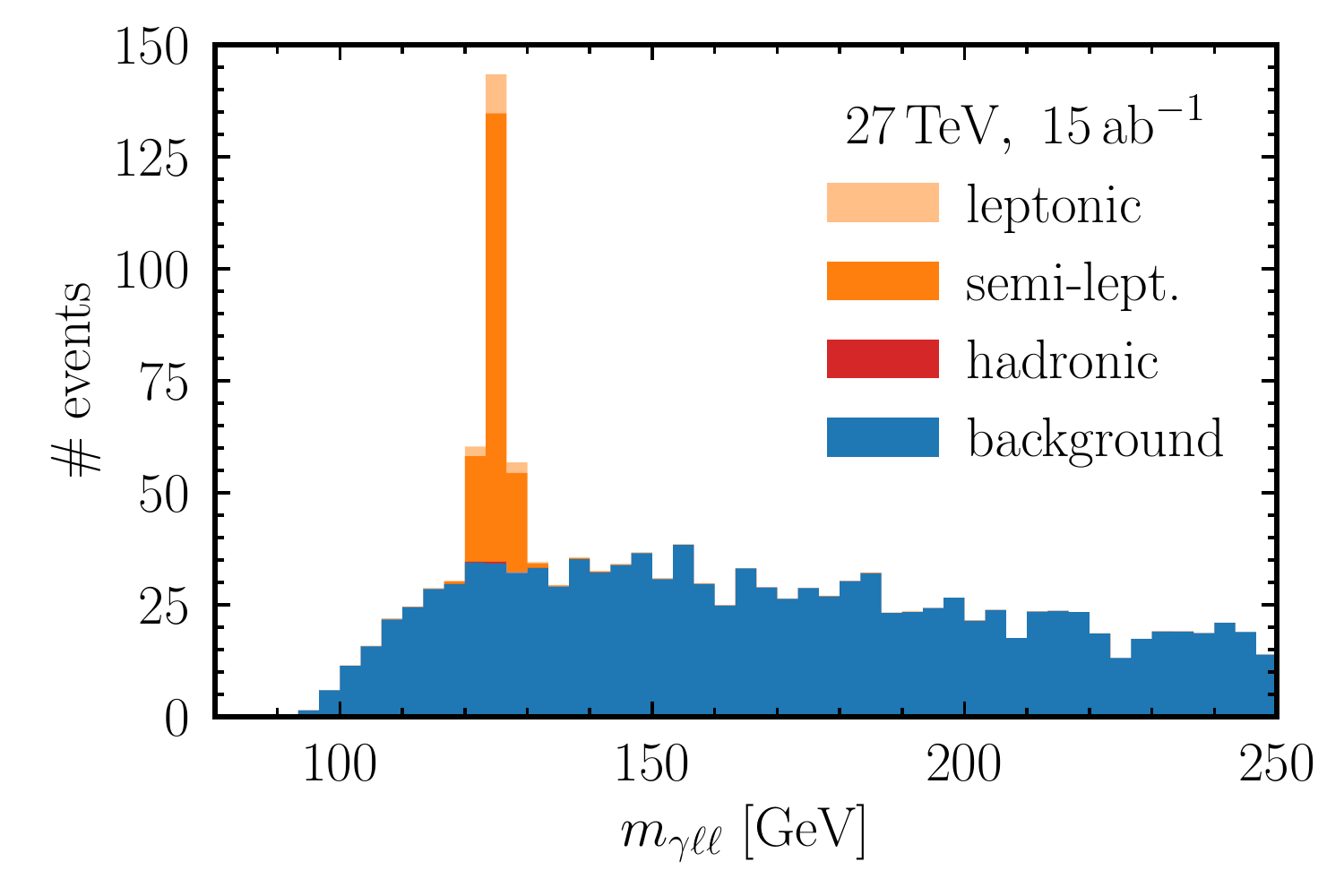}
	\includegraphics[width=.45\textwidth]{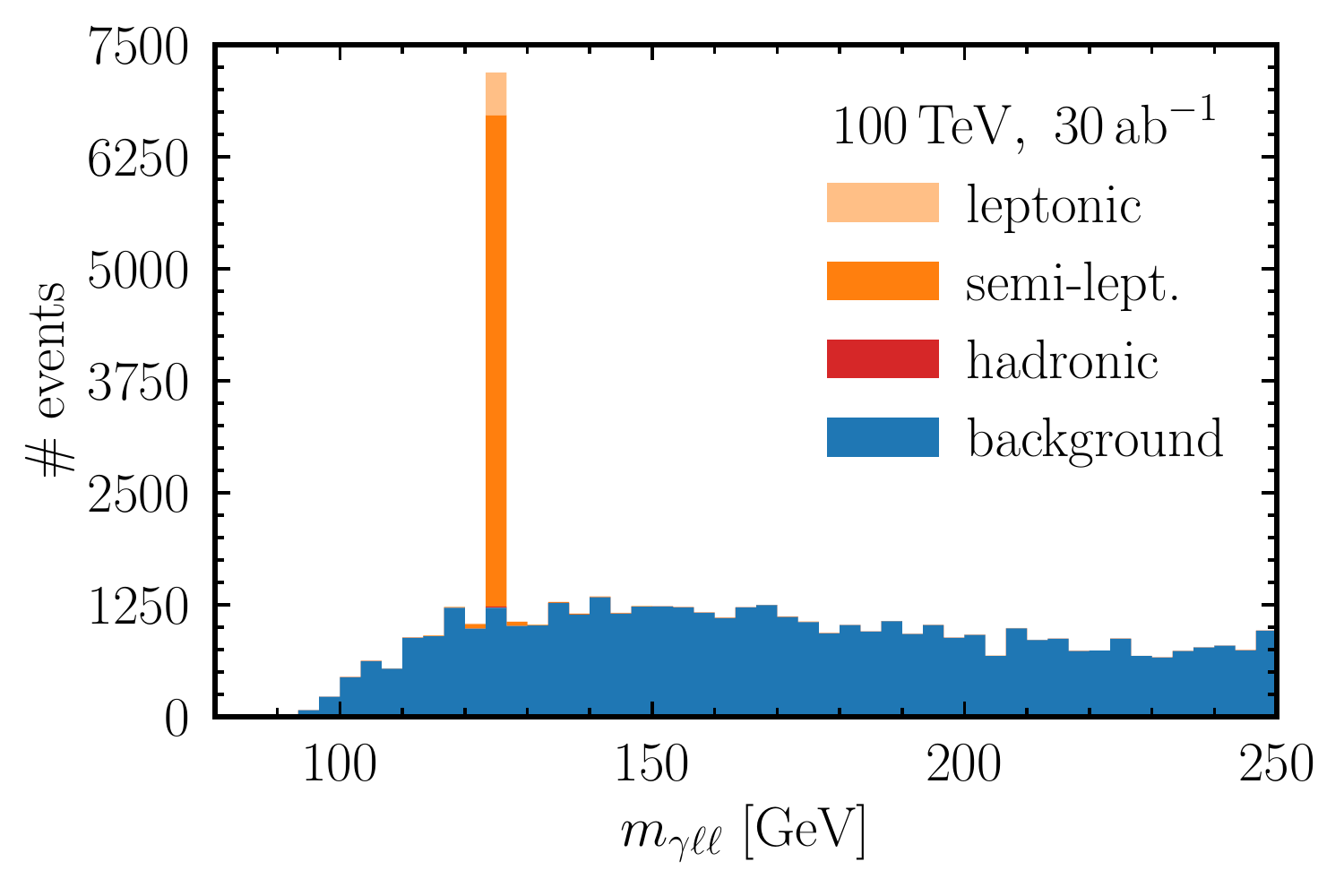}
	\caption{The invariant mass spectrum for the signal process, stacked on the background distribution (blue), before Higgs-reconstruction cut, for the top-quark pair decaying hadronically (red, not visible), semi-leptonically (orange) or leptonically (light orange).}
	\label{fig:histo27}
\end{figure*}

The invariant mass spectra of the $\ell^+\ell^-\gamma$ system at a 27~TeV and 100~TeV collider are presented in Fig.~\ref{fig:histo27}.

\begin{figure*}[!h]
	\begin{center}
		\includegraphics[width=.45\textwidth]{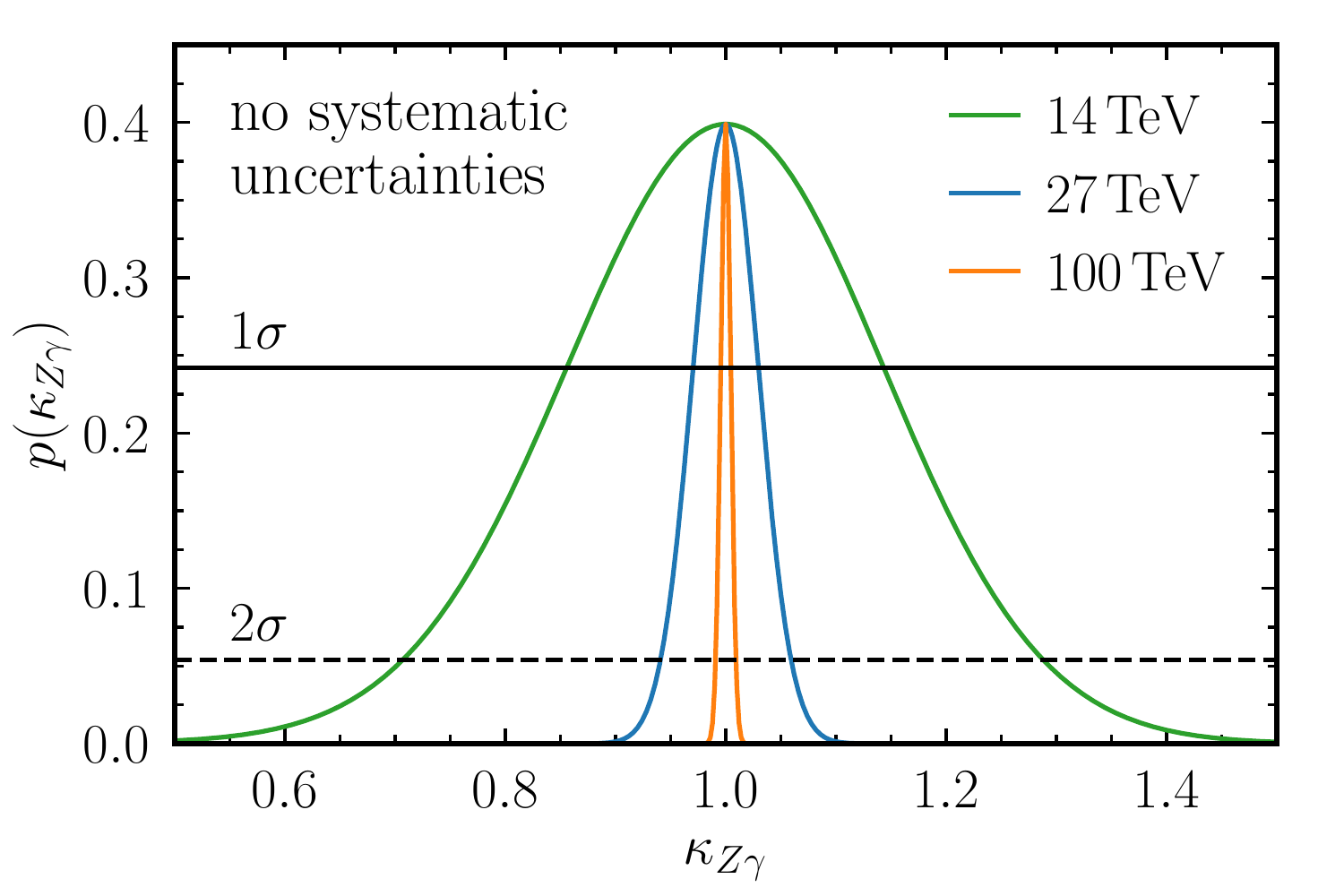}
		\includegraphics[width=.45\textwidth]{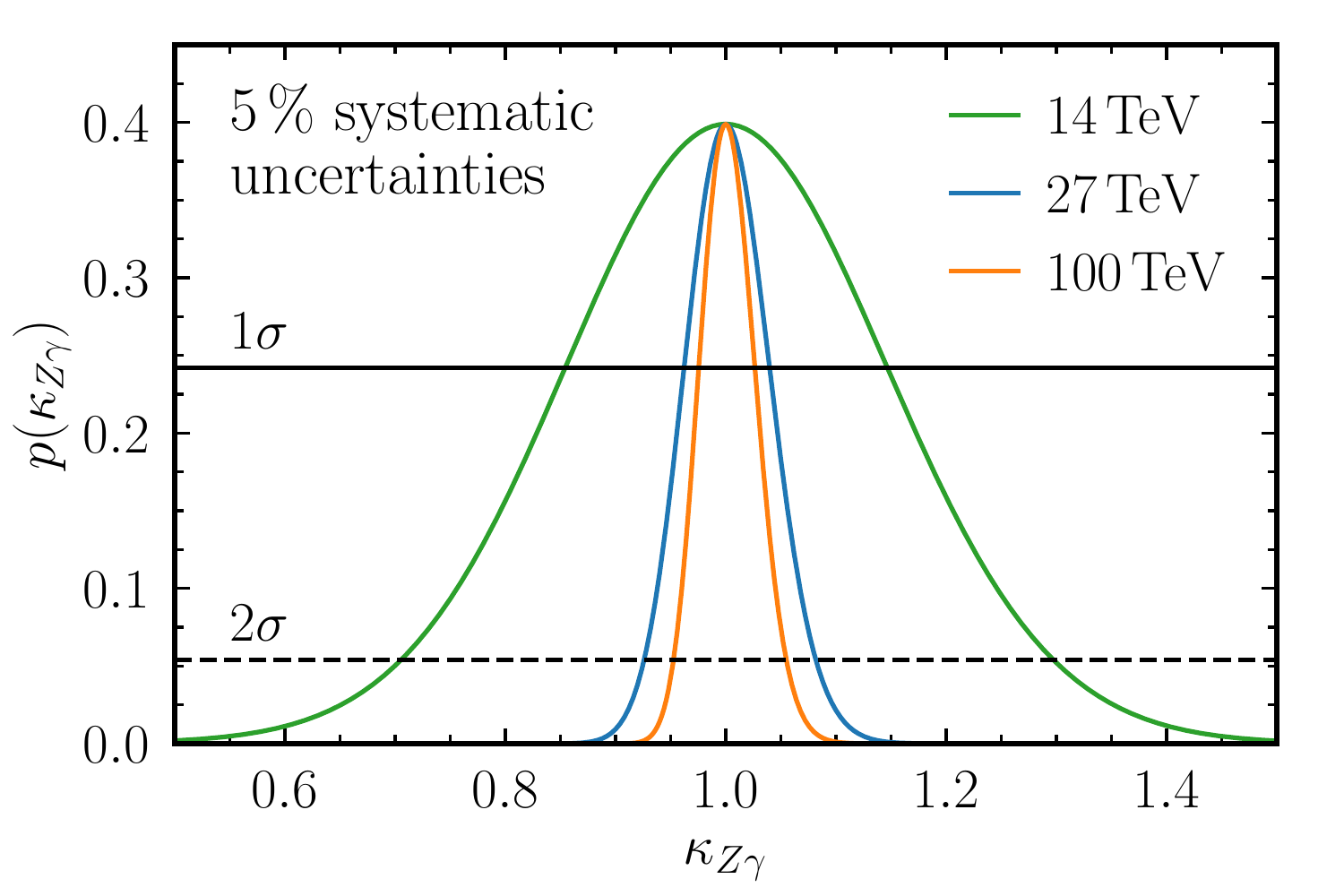}
		\caption{\label{fig:p} Left: Expected $p$-value for a given value of $\kappa_{Z \gamma}$ from the process $p p \to t \bar t h \to t \bar t Z \gamma$ at the LHC with $\sqrt{s}=14$\,TeV  and 3\,ab$^{-1}$ (green), $\sqrt{s}=27$\,TeV  and 15\,ab$^{-1}$ (blue), and FCC with $\sqrt{s}=100$\,TeV and 30\,ab$^{-1}$ (orange), assuming that the SM value is observed. Right: Same as left, but including a 5~\% systematic error. The $p$-values corresponding to $1\,\sigma$ and $2\,\sigma$ are visualised by the solid and dashed lines, respectively.
		}
	\end{center}
\end{figure*}

Furthermore in Fig.~\ref{fig:p} we show the $p$-value plots for $\kappa_{Z\gamma}$ for the 14, 27 and 100~TeV colliders, first without systematic errors and then including a 5~\% systematic error.

\end{document}